\documentclass{JHEP3}

\usepackage{graphicx}

\preprint{MIT-CTP-3595}

\title{From Entropy and Jet Quenching to Deconfinement?}

\author{Berndt M\"uller$^\dagger$ and Krishna Rajagopal$^\#$\\
~\\
$^\dagger$Department of Physics, Duke University, Durham, NC
27708-0305\\
\email{mueller@phy.duke.edu}\\
$^\#$Center for Theoretical Physics, Massachusetts Institute
of Technology,\\
Cambridge, MA 02139\\
\email{krishna@ctp.mit.edu}
}

\abstract{
The challenge of demonstrating that the matter produced
in heavy ion collisions is a deconfined quark-gluon plasma, 
as predicted by lattice QCD calculations, is the challenge
of measuring the number
of thermodynamic degrees of freedom $\nu\sim\varepsilon/T^4$
at the time $t_0$ at which 
the matter comes into approximate local thermal equilibrium
and begins to behave like a hydrodynamic fluid. Data from 
experiments done at the Relativistic Heavy Ion Collider have 
been used to estimate $t_0$ and to put a lower bound on
the energy density $\varepsilon(t_0)$. However, measuring $\nu$ has seemed out 
of reach, because no current data serve even as qualitative 
proxies for the temperature $T(t_0)$. We point out that $\nu$ 
may equally appropriately 
be defined via $\nu\sim s^4/\varepsilon^3$,
where $s$ is the entropy density, which can be estimated from 
the measured final state entropy. This estimate is based on the
testable assumption of an isentropic expansion. The observation 
of jet quenching has the potential to provide an upper bound on 
the energy density at early times. Our goal is to motivate such 
an analysis by pointing out that it would set a lower bound on 
$\nu$.
}

\begin{document}

Measurements of the final state produced in heavy ion collisions
at $\sqrt{s_\textup{NN}}=$ 130 and 200 GeV at the Relativistic
Heavy Ion Collider (RHIC) in Brookhaven have provided strong evidence
that the matter created in these reactions is rapidly thermalized
and has many properties expected from a strongly coupled
quark-gluon plasma~\cite{RHIC04}. What is lacking, however, is direct
evidence that could be linked to the prevalence of unconfined colored
excitations, which are thought to be the defining feature of this
novel state of matter. 

At present, the best evidence for deconfinement
of quarks may be that
derived from the phenomenological success of valence 
quark recombination 
models~\cite{Fries:2003vb}, 
which are based on the assumption of the existence of a thermalized
phase of unconfined ``constituent quarks'' immediately
prior to hadronization. In particular, the
quark recombination models explain the different saturation levels
and thresholds for the elliptic flow of different hadrons in terms
of a universal elliptic flow pattern at the valence quark level.
A serious limitation of this evidence is that it addresses physics 
near the quark-hadron transition, which is inherently nonperturbative 
and thus not amenable to controlled theoretical approximations. It
would be desirable to obtain evidence for the liberation
of colored degrees of freedom in the matter created at RHIC
which relates to thermalized matter present during an earlier
stage of the reaction, when the temperature is higher, 
and to observables
that allow for a more controlled theoretical description.  

In QCD with quark masses as given in nature, there is no difference 
in symmetry between an equilibrated gas of hadrons and a quark-gluon 
plasma, implying that the transition between these two regimes of 
strongly interacting matter may be a continuous, albeit possibly 
rapid, crossover~\cite{QCDPhaseDiagramReviews}. 
Indeed, this is what is predicted by {\it ab initio}
calculations of QCD thermodynamics done using the methods of lattice 
gauge theory~\cite{LatticeThermoReview}. 
Hence, an experimental demonstration of deconfinement 
cannot be seen as the answer to some ``yes/no'' question, but must 
instead involve the measurement of some physical property of the 
matter created in heavy ion collisions that can also be predicted 
by controlled theoretical calculations, and which takes on quite 
different values below and above the crossover between a hadron gas 
and a quark-gluon plasma.

Because QCD is an asymptotically free theory, there are many quantities 
whose calculation is controlled at temperatures far above the crossover,
where the effective QCD coupling $\alpha_s(T)$ becomes small.  However, 
we do not expect that collisions at RHIC can create matter in this 
asymptotic regime.  This means that in looking for quantities whose 
theoretical calculation is under control, we must ask what quantities
can be calculated by rigorous numerical methods.  Lattice calculations 
of non-thermodynamic quantities, like spectral 
functions~\cite{spectral} and viscosities~\cite{viscosity}, 
are still in their infancy and are presently
restricted to QCD in the quenched approximation.  In contrast, 
lattice 
calculations of QCD thermodynamics have reached maturity because they 
can be formulated conveniently in Euclidean quantum field theory, the 
natural arena for lattice QCD.  Lattice simulations of the QCD equation 
of state including dynamical quarks of various numbers of flavors are 
available, which permit model-independent 
conclusions~\cite{LatticeThermoReview}.

In QCD thermodynamics, there is one observable whose value changes 
by more than an order of magnitude across the crossover transition 
from hadron gas to quark-gluon plasma, namely the effective number 
of thermodynamic degrees of freedom $\nu(T)$.  A common 
definition of $\nu(T)$ is via the relation
\begin{equation}
\varepsilon(T) = \frac{\pi^2}{30} \, \nu(T)\, T^4\ ,
\label{eq:epsilon}
\end{equation}
where $\varepsilon$ and $T$ are the energy density and temperature,
respectively.  For an ideal gas of massless, noninteracting constituents, 
$\nu$ counts the number of bosonic degrees of freedom plus the number 
of fermionic degrees of freedom weighted by $7/8$.  
Eq.~(\ref{eq:epsilon}) makes the measurement of $\nu$ in heavy ion 
collisions seem a remote possibility, because there is nothing in 
the current suite of data from RHIC that is thought to serve as a 
proxy for the temperature at early times. The temperature at 
freezeout, when the matter is again hadronic, is well determined 
from many measurements of the hadronic final state.  A measurement 
of the temperatures of the quark-gluon plasma presumed to be present 
at early times is one of the goals of studies of direct photon and 
dilepton emission in heavy ion collisions but, so far, no evidence 
for thermal photon or dilepton radiation has been observed at RHIC.
Even if they will be observed eventually, thermal photons will only
yield a weighted time-average of the temperature, which may not be
sufficient to determine the function $\nu(T)$.

Here, we make the simple observation that $\nu$ can equally well be 
defined via the entropy density 
\begin{equation}
s(T)=\frac{2\pi^2}{45}\, \nu(T)\, T^3\ ,
\label{eq:entropy}
\end{equation}
and hence via
\begin{equation}
\nu(T)=\frac{1215}{128\pi^2}\,\frac{s^4}{\varepsilon^3}
      =0.96\,\frac{s^4}{\varepsilon^3}\ .
\label{eq:nudefn}
\end{equation}
For an ideal gas of massless degrees of freedom, the expressions
(\ref{eq:epsilon}), (\ref{eq:entropy}) and ({\ref{eq:nudefn})
are equivalent definitions of $\nu$.  We shall take 
(\ref{eq:nudefn}) as our definition, because in so doing we 
realize that $\nu(T)$ can be measured without measuring $T$
itself. An ideal gas of ultrarelativistic pions has $\nu=3$, 
whereas an ideal gas of noninteracting gluons and three flavors 
of massless quarks has $\nu=47.5$.  

Lattice QCD calculations show that $\nu$ increases by more than 
a factor of ten over a narrow range of temperatures centered at 
a crossover temperature $T_c=170 \pm 10$~MeV~\cite{LatticeThermoReview}.  
This rapid increase in the number of degrees of freedom is the 
direct consequence of deconfinement at high temperatures in QCD 
and, if we limit ourselves to thermodynamic observables, the 
measurement of $\nu$ is the only possible route to an experimental 
demonstration of deconfinement.

QCD does not describe an ideal gas of noninteracting quarks 
and gluons, except at infinite temperatures.  And indeed, 
there is growing evidence from the RHIC experiments that the 
matter they are creating is strongly interacting, thermalizing 
rapidly, flowing like a liquid, and opaque to energetic partons 
produced within it~\cite{RHIC04}.  
In the case of strongly interacting matter,
the different definitions of $\nu$ provided by (\ref{eq:epsilon}), 
(\ref{eq:entropy}) and ({\ref{eq:nudefn}) are not equivalent, 
and furthermore $\nu$ cannot be considered as a count of 
well-defined degrees of freedom.  Indeed, in an ideal liquid 
there are no well-defined long-lived quasiparticles.  
Nevertheless, upon choosing one definition --- and we shall 
choose (\ref{eq:nudefn}) --- $\nu$ is reliably calculable on 
the lattice, potentially measurable as we shall discuss, and 
remains a valid measure of deconfinement.

Lattice QCD 
calculations indicate that throughout the temperature
range $2 T_c < T < 5 T_c$, $\nu(T)$ is between 70\% and 80\%
of that for an ideal quark-gluon plasma~\cite{LatticeThermoReview}, 
meaning that 
$33 \lesssim \nu \lesssim 38$.  At lower temperatures, closer 
to the crossover, the value of $\nu$ extracted from lattice QCD
calculations is more significantly different for different 
definitions of $\nu$,
because just above the crossover
the pressure deviates more from its ideal gas value than
the energy density does.
At $T=1.5\,T_c$, for example, with the definition 
(\ref{eq:epsilon}) lattice calculations~\cite{LatticeThermoReview}
yield $\nu\approx 37$, whereas 
with our  
definition (\ref{eq:nudefn}) they yield $\nu\approx 27$.

\begin{figure}
\label{figdeco2}
\centerline{
\includegraphics[height=0.8\linewidth,angle=-90]{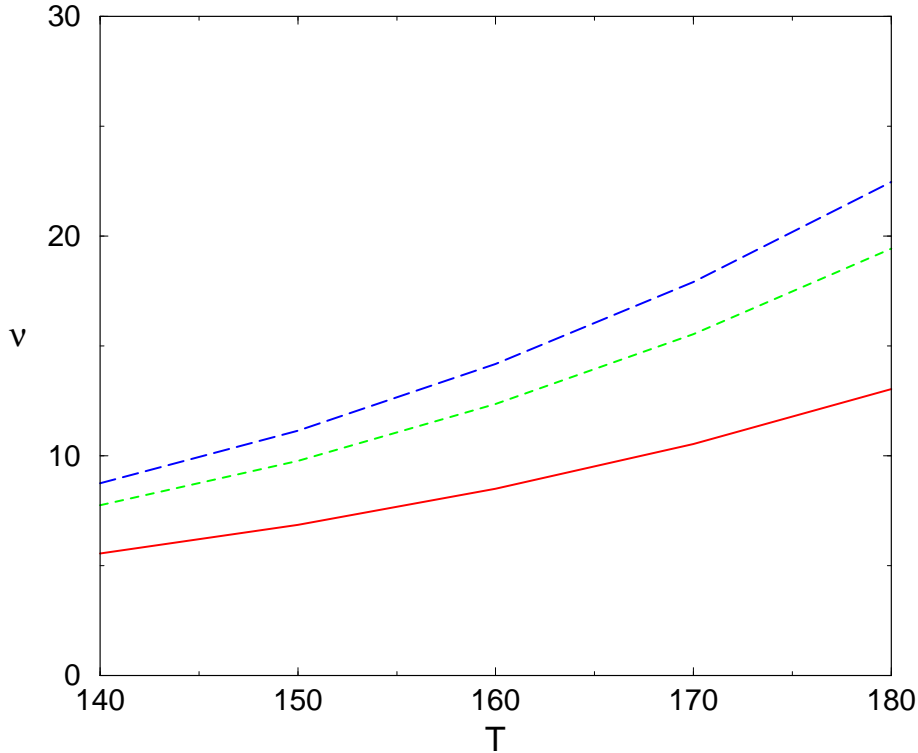}}
\caption{The effective number of degrees of freedom of a hadronic
resonance gas, 
using the three definitions of $\nu$ provided
by Eqs. (1-3): (1) long-dashed line (blue),
(2) short-dashed line (green), and (3) solid line (red). The solid
curve moves by about 5\% depending on whether we use all
established resonances in the particle data book or only
those commonly included in the chemical freezeout analysis.
This can be considered an estimate of the theoretical
uncertainty in the curve.}
\end{figure}

There is an additional benefit to defining $\nu$ via (\ref{eq:nudefn})
as we do, 
over and above making the quantity
observable in practice.  In order
to be convincing, a future experimental measurement of 
$\nu$ that agreed with what lattice QCD predicts for the 
quark-gluon plasma would need to discriminate between
this prediction and that of a hadronic resonance gas with $T\lesssim T_c$.
We see from
Fig.~1 that, with our definition of $\nu$, this requires excluding
$\nu\approx 10$.  
The three different definitions
of $\nu$ all agree for a gas of noninteracting massless 
particles, but massive particles --- as in a resonance
gas --- contribute somewhat less to $s/T^3$ than 
they do to $\varepsilon/T^4$,
and hence less still to $s^4/\varepsilon^3$.  This makes
our definition of $\nu$ advantageous.

Before turning to the challenges associated with measuring $s$ 
and $\varepsilon$, we note that the value of $\nu$ defined from 
them via (\ref{eq:nudefn}) can only be compared to QCD 
thermodynamics if there is independent evidence that the 
matter under study is in (approximate) local thermal 
equilibrium.  At RHIC, such evidence is believed to be provided 
by the agreement of the elliptic flow measured in noncentral 
collisions with hydrodynamic model predictions~\cite{RHIC04}. 
Such predictions are based on the assumption that the matter
behaves like a fluid in local thermal equilibrium,
with arbitrarily short mean free paths and
correspondingly strong interactions. The fact that the data
show as much elliptic flow as predicted indicates that this
assumption must already be valid soon after
the collision, early enough
that the azimuthal spatial anisotropy due to the nonzero 
impact parameter has not had time to be significantly reduced via free 
streaming of weakly interacting quasiparticles.  Quantitatively, it 
is estimated that local thermal 
equilibrium and the onset of hydrodynamic behavior must occur 
by a time $t_0\sim 0.6-1.0$~fm/c~\cite{Heinz:2001xi}.\footnote{ 
Recent work suggests 
that this rapid approach to 
local thermal equilibrium may occur via plasma 
instabilities, not collisions~\cite{Arnold:2004ti}.  
We caution that if thermal but not chemical equilibrium has been 
achieved at $t_0$, then $\nu(t_0)$ may
be lower than the value calculated in lattice QCD.} 
We shall take 
$t_0=1.0$~fm/c in the following.
In the remainder of this paper, we shall discuss how $s(t_0)$ 
and $\varepsilon(t_0)$ may be estimated, using present and 
near-future data.  

The estimate of $s(t_0)$ relies on the fact
that the entropy of an ideal 
fluid is conserved
during its hydrodynamical evolution.  The value of $s(t_0)$ can
therefore be deduced from the value of $dS/dy$ at
chemical freezeout, upon assuming that the fluid remains in
local thermal equilibrium between the time $t_0$ when
this condition is established and the time at which
chemical freezeout occurs. Note that we shall need
to know the volume of the system at time $t_0$, which
is early enough that little transverse expansion has taken
place making this easy to estimate, but we shall not need
to know the volume of the system at freezeout. Indeed,
we do not need to know 
anything about the
system at freezeout except its
entropy. In an isentropic expansion, it is entropy that is conserved,
regardless of how entropy density evolves.

There are, in principle, two ways of estimating $dS/dy$.
One approach uses an analysis of the composition of the
fireball at chemical freezeout to derive the entropy per hadron,
which can then be used to
relate the measured charge particle multiplicity
$dN_{\rm ch}/dy$ to the entropy $dS/dy$, assuming full thermal 
phase space occupation at the freezeout time.
The other approach~\cite{Pal:2003rz} 
uses the measured multiplicities of stable hadrons
together with experimental data on Hanbury-Brown--Twiss (HBT)
two-particle interferometry to estimate the full phase space
distributions $f_i({\vec r},{\vec p})$ 
at kinetic freezeout, and from them the entropy
\begin{equation}
S = \sum_i \int \frac{d^3r d^3p}{(2\pi)^3} 
    [-f_i \ln f_i \pm (1\pm f_i) \ln (1\pm f_i)]\ ,
\end{equation}
where the upper (lower) sign holds for bosons (fermions).

We begin with a review of the results of the kinetic freezeout
analysis. Pal and Pratt~\cite{Pal:2003rz} have used final state 
multiplicities and ``radius'' parameters deduced from a
HBT correlation analysis to determine 
$dS/dy$ in data from $\sqrt{s}=130\,A$GeV Au+Au collisions.
For collisions in a 11\% centrality window, 
Pal and 
Pratt find $dS/dy \approx 4450$ with an estimated systematic
error of $\pm 400$.  
Two effects may affect the reliability of this 
estimate. First, the conventional extraction of HBT radius
parameters is 
based on the assumption that hadrons originate from a source 
that is Gaussian in both position and momentum space.  This
is not a good assumption for a rapidly expanding source,
and the measured HBT radius parameters do not actually serve
as estimates of
the radii of the fireball at freezeout.  Hence, a more
direct measurement of the phase space distributions themselves,
without introducing a parametrization in terms of ``radii'',
would be of value. And, second, 
the entropy of the hadronic gas must increase 
as it cools after chemical freezeout but before
kinetic freezeout.  As the mean free path becomes larger
and eventually approaches the size of the system at 
kinetic freezeout, the viscosity of the hadron gas
grows, ideal hydrodynamics ceases to
be a good approximation, and entropy is produced.
Furthermore, after chemical freezeout more resonances
decay than are produced, again increasing the entropy.
Hence, the entropy at chemical freezeout, which
is what is of interest to us, must
be smaller than that estimated at kinetic freezeout.

Next, we consider the analysis based on the inferred chemical
freezeout multiplicities of hadrons, which
has the potential to extract the entropy at
chemical freezeout directly. The ratios of hadron 
multiplicities in RHIC collisions with $\sqrt{s}=200\,A$GeV are 
quite well described by the assumption that chemical freezeout 
occurs at $T_{\rm ch}=170\pm 10$~MeV from an equilibrated 
hadron gas at this temperature~\cite{Braun-Munzinger:2003zz}.
An ideal gas of all established meson
and baryon resonances at $T=170$~MeV has an 
entropy of about 7.25 per hadron~\cite{Nonaka:2004,Sollfrank:1992ru}, 
which can be compared
to the value of 3.6 for an ideal gas of massless pions.
After all the resonances 
decay the multiplicity of 
charged hadrons in the final state is 1.04 per hadron in the 
equilibrated hadron gas~\cite{Nonaka:2004}. 
In the 6\% most central collisions with $\sqrt{s}=200\,A$GeV,
the multiplicity 
of charged particles in the final state is $dN_{\rm ch}/d\eta 
= 665\pm 26$ at mid-rapidity~\cite{Back:2002uc}, corresponding to 
a $dN_{\rm ch}/dy$ that is about 10\% larger~\cite{PHOBOS}.  
Putting the pieces together, we 
estimate $dS/dy = (665 \times 1.1 \times 7.25)/1.04\approx 5100$
at chemical freezeout at mid-rapidity 
in RHIC collisions with $\sqrt{s}=200$~AGeV.  There is a 4\% uncertainty
in this estimate coming from that in 
the experimental measurement of $dN_{\rm ch}/d\eta$.
The largest theoretical uncertainty is that in
the factor $S/N=7.25$.
Changing the chemical freezeout temperature by $\pm 10$~MeV changes
$S/N$ by $\pm 3$\%. To get a sense of the other
uncertainties in $S/N$, we recalculated it using
abundances obtained including the widths of states,
and found that $S/N$ increased to 7.58.  
There are still further sources of theoretical
uncertainty that are harder to estimate like, for
example, that due to our neglect of resonances not found in the particle
data book. We do not think this is a large effect, because
reducing the number of resonances that we include does not
have a large effect.  We estimate $S/N=7.25\pm 6\%$, and expect that
a more systematic analysis could reduce this 
uncertainty by a factor of two.  Adding the theoretical and
experimental uncertainties in quadrature and rounding
upwards yields $dS/dy=5100\pm 400$.  

We can also compare our 
estimate of $dS/dy$ at chemical freezeout
to that obtained by Pal and Pratt 
at kinetic freezeout.
Applying our argument to the 11\% most central collisions
with $\sqrt{s}=130$~AGeV, for which 
$dN_{\rm ch}/d\eta=526\pm 20$~\cite{Back:2002uc},
yields the estimate 
$dS/dy = 4035\pm 300$.
As expected, the central value of this estimate
is below that obtained by Pal and Pratt,
but the difference is within error bars.
This indicates that the entropy release between
chemical and kinetic freezeout is not dramatic.  

To summarize, the best estimates available at present suggest that
$dS/dy\approx 5100\pm 400$ at chemical freezeout
at mid-rapidity 
in central RHIC collisions with $\sqrt{s}=200$~AGeV.
It seems
to us that a more careful theoretical analysis can reduce
the theoretical uncertainty by a factor of two, which would 
yield a measurement of
$dS/dy$ with 5\% errors.

To the extent that the expansion between $t_0=1$~fm/c and 
chemical freezeout is isentropic, the argument 
first proposed by Bjorken~\cite{Bjorken:1982qr} can be 
used to turn $dS/dy$ into a lower bound on $s(t_0)$. The PHOBOS 
version of this argument~\cite{PHOBOS} can be phrased as follows. 
The charged particle multiplicity is reasonably independent 
of the rapidity $y$ within the range $|y|<1$, and all this 
entropy ($S=2 dS/dy \approx 10200\pm 800 $) must have come from 
within a volume of
size $2 t_0 \pi R^2$ at time $t_0$. Taking $R=7$~fm, this yields 
$s(t_0) \ge 33\pm 3~{\rm fm}^{-3}$.  The entropy density at $t_0$ is
somewhat greater than this, because of the contribution from 
particles outside $|y|<1$.

In order to obtain a lower bound on $\nu(t_0)$ using (\ref{eq:nudefn}),
it is important to have a lower bound on $s(t_0)$, as
the Bjorken argument provides upon assuming
isentropic expansion.  It is crucial, however,
to test this assumption. That is, it is crucial to
rule out  
a significant increase of the entropy between equilibration
at time $t_0$ and chemical freezeout.  Entropy production
during the hydrodynamic expansion requires some specific mechanism
such as a strong first order phase transition which
can drive the matter away from local  thermal equilibrium.  
Lattice QCD
calculations indicate that the transition is a crossover,
with hadronization occuring continuously, 
but it would be desirable to have experimental confirmation
of the absence of a strong first order phase transition, in order
to complete the justification of our use of $dS/dy$
at chemical freezeout to obtain a lower bound on $s(t_0)$.
A strong first
order phase transition would lead to large (and possibly non-Gaussian)
event-by-event fluctuations at low $p_T$~\cite{Heiselberg:1998cv}.
There is no evidence
for such fluctuations in current data~\cite{Adams:2003st}, but
given the importance of this issue 
a more stringent 
investigation of low-$p_T$ event-by-event fluctuations is 
called for, looking at the fluctuations of several observables
and focusing on $p_T$ significantly smaller than the mean.

Just as it is important to have a lower bound on $s(t_0)$, if we 
wish to obtain a lower bound on $\nu$ we need an {\it upper} 
bound on $\varepsilon(t_0)$.  The analogous ``Bjorken argument'',
applied to $dE_T/dy$ (where $E_T$ is the total transverse energy
of the hadrons in the final state) only yields a lower bound on 
$\varepsilon(t_0)$ because the longitudinal expansion subsequent 
to $t_0$ reduces $dE_T/dy$.  For this reason, we cannot use 
$dE_T/dy$ for our purposes.  Putting this another way,
Bjorken arguments applied to $dN/dy$ and $dE_T/dy$ yield
lower bounds on both $s(t_0)$ and $\varepsilon(t_0)$, and
so by themselves these arguments give no constraint on $\nu$.

The analysis of jet quenching data has the potential to yield an 
upper bound on the energy density at early times, as we now discuss.  
High energy partons traversing strongly interacting matter lose 
energy mainly by gluon radiation after interactions with colored 
constituents of the medium. The theory of this mechanism is well 
developed within the framework of perturbative QCD 
\cite{Wang:1994fx,Baier:1996kr,Baier:1996sk,Zakharov:1996fv,Gyulassy:2000fs,Baier:2000mf}.
In the 
multiple soft scattering limit, the effect of the medium is
encoded in the parameter
\begin{equation}
{\hat q} = \rho \int q^2 dq^2 (d\sigma/dq^2) \, ,
\label{q-hat}
\end{equation}
where $\rho$ is the density of scattering centers in the medium, 
$q^2$ denotes the momentum transfer in scattering, and $d\sigma/dq^2$
is the differential cross section for scattering of the hard parton
on a single center. Coherence effects suppress the emission of
gluons with energies above $\omega_c = {\hat q} L^2 /2$, where
$L$ denotes the length of material traversed. The mean energy
lost by the parton is given by 
$\Delta E \approx (3\alpha_s C_R/\pi) \omega_c$,
where $C_R$ is the Casimir operator for the color representation
of the parton. Similar results are obtained in the opposite (low
opacity) limit, when the interaction with the medium is dominated
by a single or few scatterings~\cite{Gyulassy:2000fs}.
In these analyses, the properties of the matter being probed arise 
only in the transport coefficient $\hat q$.  In this sense, jet 
quenching can be thought of as ``measuring $\hat q$.''

The relationship between $\hat q$ and the energy density $\varepsilon$
has not been determined in general, but it is known for the limiting
cases of cold nuclear matter~\cite{Baier:1996sk}, an ideal pion 
gas, and an ideal weakly-interacting quark-gluon 
plasma~\cite{Baier:2002tc}.  When a high energy parton penetrates 
strongly interacting matter, it resolves the partonic constituents 
of the medium. In the case of cold nuclear matter, it mainly interacts 
with the gluon component of the nucleons, and the parameter $\hat q$
can be expressed in terms of the gluon distribution in the 
nucleon~\cite{Baier:1996sk}.
For a medium such as a weakly interacting quark-gluon plasma, in which 
all partons are deconfined, each quark and gluon contributes to the 
density of scatterers independently, and $\hat q$ is directly given
by Eq.~(\ref{q-hat}) in terms of the gluon density and perturbatively
screened parton-parton cross section. The resulting energy loss in
a thermal quark-gluon plasma has been calculated by Baier as function
of the energy density of the plasma \cite{Baier:2002tc}. 
Remarkably, when the energy loss coefficient is calculated for a 
thermal gas of pions, 
one finds the same
value of $\hat q$ as for a thermal quark-gluon plasma with
the same energy density.  With the benefit of
hindsight, this is not entirely 
surprising, although the precision of the agreement may be 
coincidental.  About half of the momentum of a fast moving hadron 
is carried by gluons, 
and roughly half of the energy density
of a weakly interacting quark-gluon plasma is contained in gluons.
Normalized to the energy density of the medium, a fast moving 
parton can therefore be expected to encounter roughly the same number of
gluons on which it can scatter. 

It is presumably naive to think that a measurement of $\hat q$ 
{\em is} a measurement of the energy density
$\varepsilon$, as the comparison between 
Baier's results for energy loss in a weakly interacting pion gas 
and in a weakly interacting quarki-gluon plasma would suggest.  
Certainly, the relationship between the energy loss of a hard 
parton traversing a medium and the energy density of the medium 
requires further elucidation and generalization.  The perturbative 
expression for the radiative energy loss of an energetic parton, 
which is only known to leading order in the strong coupling 
$\alpha_s$, may get substantial corrections at higher order. 
A calculation of next-to-leading order corrections to the energy
loss in perturbative QCD would be desirable. Although the energy
of the penetrating parton provides a large scale, the momentum
transfer of the scattering in the medium provides a second, much
lower energy scale, which could enter into the NLO corrections.
Also, the quantity $\hat q$ which provides the link between the 
observable $\Delta E$ and the energy density $\varepsilon$ is 
defined only in the context of the perturbative analysis. Since 
$\varepsilon$ itself is nonperturbatively well-defined, it would 
ultimately be desirable to understand the relation between parton 
energy loss observables and $\varepsilon$ directly, possibly involving other 
nonperturbatively defined properties of the medium being probed 
by the hard parton.
Keeping these caveats in mind, we nevertheless expect that the 
qualitative lessons encoded in Baier's results will survive in a 
more rigorous treatment.  

Jet quenching by itself can never provide 
a measure of $\nu$.  It cannot differentiate between a weakly 
interacting quark-gluon plasma and a system whose gluon content 
is the same, but which has much less entropy density because
the gluons are bound within hadrons and hence do not directly
contribute to the entropy.  In a hadron gas, the entropy is 
(roughly) a count of the hadrons, which are the thermodynamically 
independent degrees of freedom, whereas a hard parton ``sees'' the 
gluons within each hadron.  
A weakly interacting quark-gluon plasma 
and a hypothetical weakly interacting pion gas with the same energy 
density are equally effective at quenching jets, according to 
Baier's perturbative analysis, but the pion gas has a much lower 
entropy density and hence a lower $\nu$.  Turning it around, if 
we imagine a hadron gas and a quark-gluon plasma with the same 
entropy density, the quark-gluon plasma has the smaller energy 
density, the smaller density of gluonic scatterers (only one
per entropically active degree of freedom), and hence the 
smaller energy loss by a factor that scales like $\nu^{1/3}$.
The details of this calculation will differ depending on the
precise nature of the composition of the medium, but the principle
is general: Any hadronic medium will contain several gluons per 
hadron, but only the entire hadrons will contribute to the entropy.
For a deconfined medium, on the other hand, each gluon contributes
to the entropy individually, implying a smaller number of gluons 
(per unit of entropy) capable of scattering an energetic 
parton.\footnote{As 
an aside, we should note that one proposed way of understanding
the lattice QCD result that $\nu$ in the quark-gluon plasma
is somewhat less than that for
an ideal quark-gluon plasma invokes the possibility that,
in the energy density range relevant to the RHIC experiments, 
matter in the deconfined phase may contain 
colored bound states of quarks
and gluons \cite{Shuryak:2003ty}. This reduces the
entropy density somewhat 
without reducing the density of gluonic scatterers
in the medium.  This is another illustration of our logic, but
it is not directly
relevant to the strategy for measuring 
$\nu$ that we propose. Any bound on $\nu$ extracted from
data can be compared directly to lattice QCD calculations,
regardless of the mechanism
by which the strong interactions create the
deviation of $\nu$ from its ideal quark-gluon plasma value.
Note also that 
colored bound state formation among quarks and gluons introduces
a new contribution to the energy loss, that coming from ionization
of the bound states~\cite{Shuryak:2004ax}.  
Neglecting this contribution in a
future analysis of jet quenching would mean that the inferred
energy density is greater than the true energy density, and thus
would not interfere with the goal of obtaining an upper bound
on the energy density.}

Given a lower bound on $s(t_0)$, in order to obtain a lower bound
on $\nu$, and hence perhaps a demonstration of deconfinement,
we will need to use jet quenching data to obtain
an upper bound on $\varepsilon(t_0)$.   Further theoretical
work is required before this can be attempted quantitatively.
Even if the
relation between jet quenching data and energy density were
in hand, current data may not be sufficient to provide
an upper bound.  The fact that the observed jet quenching in
the most central collisions is quite large (the suppression 
ratio $R_{AA} \ll 1$) means that it may be difficult to derive
an upper bound on $\hat q$ from these data~\cite{Eskola:2004cr}.  
The observed
strong suppression
implies that the observed hadrons are the leading partons from 
jets originating near the surface of the matter~\cite{RHIC04}, as is also 
indicated by the absence (within current error bars) of any
high-$p_T$ hadrons from the 
away-side jet in the most central collisions~\cite{STAR}.  
Indeed, the currently ongoing careful study of the material recoiling
against a jet originating from near the surface of the matter,
for example via dihadron distributions, will provide us with
data on new jet quenching observables and could teach
us more about the matter produced in central collisions.
However, since an upper 
bound on the energy density must take the form of a statement 
that if $\varepsilon$ were larger more suppression would 
have been seen, jet quenching data from noncentral collisions
may be a more powerful source of information.  Indeed, the
away-side jet does not fully disappear in noncentral collisions, 
especially when the jets only have to fight their way through 
the hot matter in the narrow direction~\cite{STAR}, giving hope that a 
quantitative upper limit for the value of $\hat q$ may not be 
far away. Data at higher $p_T$ and with higher statistics is 
required, and should come from the Run-4 data set now being 
analyzed.  

Although the above paragraph is our
true conclusion, we would be remiss to end without
attempting to ``plug in numbers just for fun'', 
even absent a reliable upper bound on $\varepsilon(t_0)$.
By applying the Bjorken argument
to $dE_T/dy$, both PHOBOS and PHENIX 
estimate that $\varepsilon(1\,{\rm fm}/c))> 5~{\rm GeV/fm}^{-3}$~\cite{RHIC04}.
This lower bound on the energy density is of considerable interest in and 
of itself, even though by itself it cannot be used to constrain $\nu$. 
Indeed, if we take the lattice QCD calculation of 
$\nu(T)$ as a given, the experimental lower
bound on $\varepsilon$ tells us that at $t=1~{\rm fm}/c$ the matter 
produced at RHIC is a quark-gluon plasma with a temperature
$T > 1.4 \,T_c$, well above the crossover.
In other words, using (\ref{eq:epsilon}), a 
lower bound on $\varepsilon$, and the lattice calculation of $\nu$, 
we obtain a lower bound on $T$ and a ``demonstration'' of 
deconfinement, albeit one that is
unsatisfying because it uses the lattice calculation 
of $\nu$ rather than testing it.  
In order to measure $\nu$ and (presumably) demonstrate
deconfinement,
what is required is a lower bound on $s(t_0)$ and 
an upper bound on $\varepsilon(t_0)$.
If we adopt the conclusion
from above that $s(t_0)> 33\pm 3~{\rm fm}^{-3}$ and,
absent a reliable upper bound, 
suppose that $\varepsilon(1\,{\rm fm}/c)$ is given
by  5, 7 or 9~${\rm GeV/fm}^{-3}$,
we would conclude that $\nu>71\pm 22$, 
$\nu>26\pm 8$ or $\nu>12\pm 4$, respectively.
A 5\% determination of $s(t_0)$ would reduce these
error bars significantly,
which motivates the theoretical effort needed
to accomplish this goal. Another direction
in which theoretical effort is needed is the
modelling of the consequences 
of the variation of $s$ and $\varepsilon$ across
the transverse extent of the collision region,
something we have not considered here.
A stringent experimental investigation of low-$p_T$ 
event-by-event fluctuations is also required, in order to augment
current theoretical evidence with experimental
evidence against a strong first
order phase transition, whose attendant entropy production
would complicate the extraction of $\nu$ that we propose.
And, most important, these numbers make very clear
the importance of further analysis of jet quenching
theory and data with the goal of setting a reliable
upper bound on $\varepsilon(t_0)$.  If there were
experimental evidence that $\varepsilon(1\,{\rm fm}/c)<7\,{\rm GeV/fm}^{-3}$,
this would be evidence for deconfinement.  

Looking further ahead, if RHIC data can provide
interesting limits on the value of $\nu$,
data from heavy ion collisions at the LHC should do even
better.   And, QCD predicts that if $\nu$ 
is above the crossover at RHIC, its value will not
increase significantly at the LHC.  It is a greater
challenge to devise a way of measuring $\nu$ at lower
energy densities using lower energy heavy ion collisions,
where jet quenching is not observable.

\vfill\eject

\parindent=0in
{\bf Acknowledgments:} 
We thank R. Baier, W. Busza, U. Heinz, 
P. Jacobs, T. Renk, G. Roland, X. Wang, U. Wiedemann and W. Zajc for
helpful discussions. We also thank the organizers 
and participants of the 
RIKEN-BNL workshop ``New
Discoveries at RHIC'', which provided the stimulus
for this work, and the organizers and participants
of the ``Hard Probes 2004'' conference in Ericeira, Portugal, where this
work was discussed.
This research was supported in part by the U.S.\
Department of Energy (D.O.E.) under grant \#DE-FG02-96ER40945
and cooperative research agreement \#DF-FC02-94ER40818.


\begin{thebibliography}{99}

\bibitem{RHIC04}
I. Arsene {\it et al.} [BRAHMS Collaboration], arXiv: nucl-ex/0410020;
K. Adcox {\it et al.} [PHENIX Collaboration], arXiv:nucl-ex/0410003;
B. B. Back {\it et al.} [PHOBOS Collaboration], arXiv: nucl-ex/0410022;
J. Adams {\it et al.} [STAR Collaboration], arXiv: nucl-ex/0501009.

\bibitem{Fries:2003vb}
R.~J.~Fries, B.~M\"uller, C.~Nonaka and S.~A.~Bass,
Phys.\ Rev.\ Lett.\  {\bf 90}, 202303 (2003)
[arXiv:nucl-th/0301087];
Phys.\ Rev.\ C {\bf 68}, 044902 (2003)
[arXiv:nucl-th/0306027];
%
V.~Greco, C.~M.~Ko and P.~Levai,
Phys.\ Rev.\ Lett.\  {\bf 90}, 202302 (2003)
[arXiv:nucl-th/0301093];
Phys.\ Rev.\ C {\bf 68}, 034904 (2003)
[arXiv:nucl-th/0305024];
%
D.~Molnar and S.~A.~Voloshin,
Phys.\ Rev.\ Lett.\  {\bf 91}, 092301 (2003)
[arXiv:nucl-th/0302014];
%
R.~C.~Hwa and C.~B.~Yang,
Phys. Rev. C {\bf 70}, 024905 (2004)
[arXiv:nucl-th/0401001];
%
R.~J.~Fries,
J.\ Phys.\ G {\bf 30}, S853 (2004)
[arXiv:nucl-th/0403036].

\bibitem{QCDPhaseDiagramReviews}
For reviews, see K. Rajagopal, arXiv:hep-ph/9504310;
Acta Phys. Polon. B {\bf 31}, 3021 (2000)
[arXiv:hep-ph/0009058].

\bibitem{LatticeThermoReview}
For reviews, see 
F.~Karsch,
Nucl.\ Phys.\ A {\bf 698}, 199 (2002)
[arXiv:hep-ph/0103314];
Lect. Notes Phys. {\bf 583}, 209 (2002) 
[arXiv:hep-lat/0106019];
P. Petreczky, arXiv:hep-lat/0409139.



\bibitem{spectral}
F.~Karsch,
J.\ Phys.\ G {\bf 30}, S887 (2004)
[arXiv:hep-lat/0403016].
M.~Asakawa and T.~Hatsuda,
Nucl.\ Phys.\ A {\bf 721}, 869 (2003).

\bibitem{viscosity}
F. Karsch and H. W. Wyld, Phys. Rev. D {\bf 35}, 2518 (1987);
G.~Aarts and J.~M.~Martinez Resco,
JHEP {\bf 0204}, 053 (2002)
[arXiv:hep-ph/0203177];
S. Gupta, Phys. Lett. B {\bf 597}, 57 (2004) [arXiv:hep-lat/0301006];
A. Nakamura and S. Sakai, arXiv:hep-lat/0406009.

\bibitem{Heinz:2001xi}
U.~W.~Heinz and P.~F.~Kolb,
Nucl.\ Phys.\ A {\bf 702}, 269 (2002)
[arXiv:hep-ph/0111075].


\bibitem{Arnold:2004ti}
P. Romatschke and M. Strickland, Phys. Rev. D {\bf 68}, 036004 (2004)
[arXiv:hep-ph/0304092];
P. Arnold, J. Lenaghan and G. D. Moore, JHEP {\bf 0308}, 002 (2003)
[arXiv:hep-ph/0307325];
P.~Arnold, J.~Lenaghan, G.~D.~Moore and L.~G.~Yaffe,
arXiv:nucl-th/0409068;
P.~Romatschke and M.~Strickland,
arXiv:hep-ph/0408314.

\bibitem{Pal:2003rz}
S.~Pal and S.~Pratt,
Phys.\ Lett.\ B {\bf 578}, 310 (2004)
[arXiv:nucl-th/0308077].



\bibitem{Braun-Munzinger:2003zz}
P.~Braun-Munzinger, J.~Stachel and C.~Wetterich,
Phys.\ Lett.\ B {\bf 596}, 61 (2004)
[arXiv:nucl-th/0311005];
M.~Kaneta and N.~Xu,
arXiv:nucl-th/0405068;
for a review, see P. Braun-Munzinger, K. Redlich and J. Stachel,
arXiv:nucl-th/0304013.

\bibitem{Nonaka:2004}
C.~Nonaka, S.~A.~Bass, B.~M\"uller, and M.~Asakawa,
arXiv:nucl-th/0501028.

\bibitem{Sollfrank:1992ru}
For earlier work in this direction, see
J. Sollfrank and U. W. Heinz,  Phys. Lett. B {\bf 289}, 132 (1992).


\bibitem{Back:2002uc}
B.~B.~Back {\it et al.} [PHOBOS Collaboration] 
Phys.\ Rev.\ C {\bf 65}, 061901 (2002) [arXiv:nucl-ex/0201005].

\bibitem{PHOBOS}
B. B. Back {\it et al.} in Ref.~\cite{RHIC04}, and references
therein.

\bibitem{Bjorken:1982qr}
J.~D.~Bjorken,
Phys.\ Rev.\ D {\bf 27}, 140 (1983).

\bibitem{Heiselberg:1998cv}
H.~Heiselberg and A.~D.~Jackson,
arXiv:nucl-th/9809013;
Phys.\ Rev.\ C {\bf 63}, 064904 (2001)
[arXiv:nucl-th/0006021];
I.~N.~Mishustin,
Phys.\ Rev.\ Lett.\  {\bf 82}, 4779 (1999)
[arXiv:hep-ph/9811307].



\bibitem{Adams:2003st}
K. Adcox {\it et al.} [PHENIX Collaboration], 
Phys. Rev. C {\bf 66}, 024901 (2002) 
[arXiv:nucl-ex/0203015];
J.~Adams {\it et al.} [STAR Collaboration],
Phys.\ Rev.\ C {\bf 68}, 044905 (2003)
[arXiv:nucl-ex/0307007];
arXiv:nucl-ex/0308033;
S. S. Adler {\it et al.} [PHENIX Collaboration], 
Phys. Rev. Lett. {\bf 93}, 092301 (2004) [arXiv:nucl-ex/0310005].

\bibitem{Wang:1994fx}
X.~N.~Wang, M.~Gyulassy and M.~Plumer,
Phys.\ Rev.\ D {\bf 51}, 3436 (1995)
[arXiv:hep-ph/9408344].

\bibitem{Baier:1996kr}
R.~Baier, Y.~L.~Dokshitzer, A.~H.~Mueller, S.~Peigne and D.~Schiff,
Nucl.\ Phys.\ B {\bf 483}, 291 (1997)
[arXiv:hep-ph/9607355].

\bibitem{Baier:1996sk}
R.~Baier, Y.~L.~Dokshitzer, A.~H.~Mueller, S.~Peigne and D.~Schiff,
Nucl.\ Phys.\ B {\bf 484}, 265 (1997)
[arXiv:hep-ph/9608322].

\bibitem{Zakharov:1996fv}
B.~G.~Zakharov,
JETP Lett.\  {\bf 63}, 952 (1996)
[arXiv:hep-ph/9607440];
R.~Baier, Y.~L.~Dokshitzer, A.~H.~Mueller and D.~Schiff,
Phys.\ Rev.\ C {\bf 58}, 1706 (1998)
[arXiv:hep-ph/9803473];
X.~N.~Wang and X.~F.~Guo,
Nucl.\ Phys.\ A {\bf 696}, 788 (2001)
[arXiv:hep-ph/0102230];
E.~Wang and X.~N.~Wang,
Phys.\ Rev.\ Lett.\  {\bf 89}, 162301 (2002)
[arXiv:hep-ph/0202105].

\bibitem{Gyulassy:2000fs}
M.~Gyulassy, P.~Levai and I.~Vitev,
Phys.\ Rev.\ Lett.\  {\bf 85}, 5535 (2000)
[arXiv:nucl-th/0005032];
Nucl.\ Phys.\ B {\bf 594}, 371 (2001)
[arXiv:nucl-th/0006010];
U. A. Wiedemann, Nucl. Phys. B {\bf 588}, 303 (2000) [arXiv:hep-ph/0005129].

\bibitem{Baier:2000mf}
For reviews, see 
R.~Baier, D.~Schiff and B.~G.~Zakharov,
Ann.\ Rev.\ Nucl.\ Part.\ Sci.\  {\bf 50}, 37 (2000)
[arXiv:hep-ph/0002198];
A. Kovner and U. A. Wiedemann, in {\it Quark Gluon Plasma 3}, eds.
R. C. Hwa and X. N. Wang (World Scientific, Singapore, 2004) 
[arXiv:hep-ph/0304151];
A.~Accardi {\it et al.},
arXiv:hep-ph/0310274.

\bibitem{Baier:2002tc}
R.~Baier,
Nucl.\ Phys.\ A {\bf 715}, 209 (2003)
[arXiv:hep-ph/0209038].


\bibitem{Shuryak:2003ty}
E.~V.~Shuryak and I.~Zahed,
Phys.\ Rev.\ C {\bf 70}, 021902 (2004)
[arXiv:hep-ph/0307267];
%
Phys. Rev. D {\bf 70}, 054507 (2004)
[arXiv:hep-ph/0403127].

\bibitem{Shuryak:2004ax}
E.~V.~Shuryak and I.~Zahed,
arXiv:hep-ph/0406100.

\bibitem{Eskola:2004cr}
K.~J.~Eskola, H.~Honkanen, C.~A.~Salgado and U.~A.~Wiedemann,
Nucl.\ Phys.\ A {\bf 747}, 511 (2005)
[arXiv:hep-ph/0406319].


\bibitem{STAR} 
C. Adler {\it et al.} [STAR Collaboration], Phys. Rev. Lett. {\bf 90},
082302 (2003);
J. Adams {\it et al.} [STAR Collaboration], Phys. Rev. Lett. {\bf 91},
072304 (2003); Phys. Rev. Lett. {\bf 93},
252301 (2004) [arXiv:nucl-ex/0407007];
J. Adams {\it et al.} in Ref.~\cite{RHIC04}.


\end{thebibliography}
\end{document}